# Optoacoustic flow-cytometry with light scattering referencing

**Markus Seeger[a,b], Andre C. Stiel[b*] and Vasilis Ntziachristos[a,b*]**

[a]Chair of Biological Imaging and Center for Translational Cancer Research (TranslaTUM) Technische Universität München, Munich, Germany

[b]Institute of Biological and Medical Imaging (IBMI), Helmholtz Zentrum München, Neuherberg, Germany

[*]Vasilis.ntziachristos@helmholtz-muenchen.de or andre.stiel@helmholtz-muenchen.de

**Abstract**

In analogy to the development of fluorescent proteins, innovative tools for screening optoacoustic cell labels could lead to tailored protein labels for OA, imparting novel ways to visualize biological structure and function. Optoacoustic imaging emerges towards a highly promising modality for life sciences and medical practise with advantageous capabilities such as great accessible depth, and 3D studying of living tissue. The development of novel labels with molecular specificity could significantly enhance the optoacoustic contrast, specificity, and sensitivity and allow optoacoustic to interrogate tissues not amenable to the fluorescence method. We report on an optoacoustic flow cytometer (OAFC) prototype, developed for screening optoacoustic reporter genes. The cytometer concurrently records light scattering for referencing purposes. Since recording light scattering is completely independent from OA, we believe it to be a more reliable referencing method than e.g. fluorescence or ultrasound-backscatter. Precise characterization of our OAFC prototype showcases its ability to optoacoustically characterize objects in-flow that are in the size range of single cells. We apply the OAFC to distinguish individual *E. coli* cells based on optoacoustic properties of their expressed chromoproteins read in-flow using microfluidic arrangements and achieved precisions over 90%. We discuss how the light scattering referenced OAFC method offers a critical step towards routine measurement of optoacoustic properties of single-cells and could pave the way for identifying genetically encoded optoacoustic reporters, by transferring working concepts of the fluorescence field.

**Abbreviations**

| | |
|---|---|
| AUC | Area under curve |
| CNP | Carbon nano particles |
| *E. coli* | *Escherichia coli* |
| FN | False negative |
| LS | Light scattering |
| µB | microbeads |
| mKO | mKusabira Orange1 |
| OA | Optoacoustic |
| OAFC | Optoacoustic flow-cytometer |
| ROC | Receiver operator characteristics |
| TP | true positive |

**Introduction**

High-throughput analysis of cells based on their fluorescence characteristics is by now a standard laboratory tool enabling numerous medical applications as well as routine analysis and sorting of cells for life science applications. Used in techniques such as fluorescence-activated cell sorting (FACS), flow cytometry reads fluorescent dyes or genetically encoded fluorescent proteins in single-cells, which allows its employment for the development of novel contrast agents as well as it's usage for profiling a population of cells. Nevertheless, since fluorescence-based flow cytometers rely on extracting information of the cells via fluorescence labels, such studies are restricted in both 1) only impacting the field of fluorescence-based imaging and sensing, and 2) information accessible that is addressable via fluorescence labels.

Next to fluorescence as a modality to investigate biological specimen, photo- or optoacoustic (OA) imaging currently emerges towards a highly promising imaging and sensing methodology in biomedical science and clinical practise. Relying on optical absorption but reading ultrasound generated from non-radiative energy dissipation, OA imaging overcomes limitations of fluorescence-based methods by a greater accessible depth in biological tissue, potentially label-free contrast, intrinsic 3D measurements, high spatial and temporal scalability, and vast potential for translating findings of basic research to the clinics. Transferring the concept of high-throughput fluorescence-based analysis of single-cells to the OA field would allow the implementation of directed evolution techniques to develop novel genetically encoded OA-contrast agents. However, accurate high-throughput in-flow OA measurement of intracellular contrast is challenging as the recorded OA signals are strongly affected by the cell's position in the interrogation area, the cell's size, and the sensitivity of the system. This hurdle renders the need for appropriately referencing the detected OA signals in order to account for the above-mentioned factors.

OA flow cytometry has so far been implemented for *in vivo* measurements (extensively reviewed in Galanzha et al.[1] and van den Berg et al.[2]) and to a limited extent for *in vitro* measurements[3–6]. However, published *in vitro* OA flow cytometry approaches rely on recording fluorescence or ultrasound-backscatter signals simultaneously with OA used for normalization and referencing[7,8]. In the case of the former, OA and fluorescence signals originate from the same underlying absorption process, they are not independent and, thus, do not represent a valuable strategy for signal normalization. In the case of the latter, ultrasound is characterized typically with a much poorer spatial resolution, which thereby prevents accurate correction or normalization of the corresponding OA signal.

To address this issue of appropriate referencing the OA signal, here we present the prototype of combined in-flow detection of OA and detection of light-scattering (LS). The key hypothesis was to account for the cell's properties inferred from the LS for referencing the OA signal. The LS signal examines herein the overall sensed volume per cell and, thus, can be used for normalizing cell size and position. It therefore represents a perpendicular methodology to OA readouts. Here we present the pilot of an OA flow cytometer (OAFC) relying on LS referencing towards providing robust routine measurement of cellular absorption profiles.

**Results**

The designed OAFC (Fig. 1a-b and methods section) consists of a 50 kHz pulsed 532 nm laser excitation focused on a standard 1000 μm x 100 μm microfluidic chip through which the cells are guided; resulting ultrasound signals are recorded by a 50 MHz focused transducer in perpendicular orientation, while LS signals are detected by a fiber-coupled photodiode in an orientation 45° angled to the OA detection.

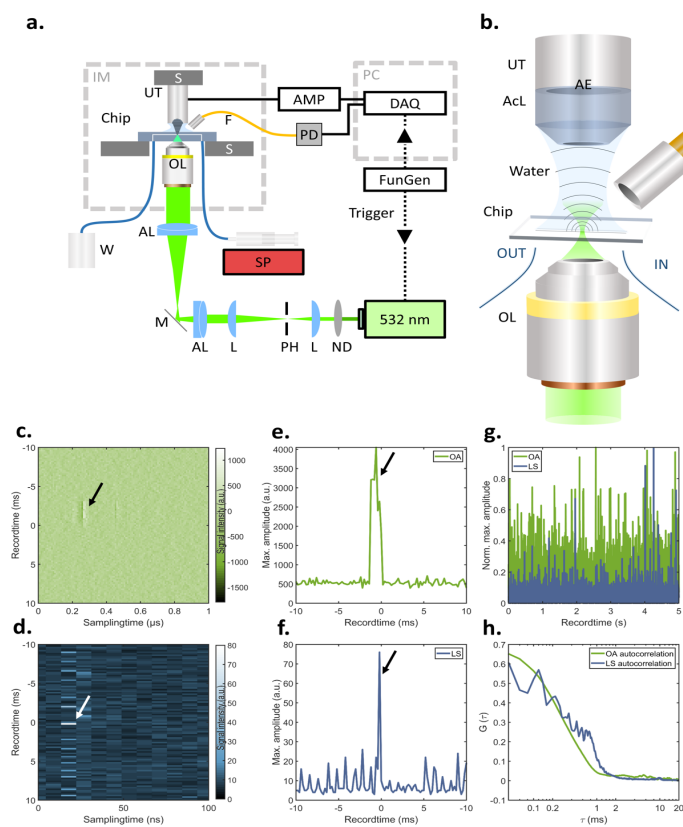

*Fig. 1 Schematic depiction and validation of the OAFC system. (a) The OAFC system is based on focusing a 532 nm laser into a microfluidic channel, which is connected to a syringe pump for controlling the flow of suspensions of particels or cells. OA and LS signals are recording simultaneously. (b) Close-up depiction of the microfluidic interrogation area. Raw (c) OA and (d) LS signal detected from a single bead indicated by an arrow flowing through the interrogation area flowing at 100 μL/min. (e) OA and (f) LS maximum amplitude projection of the record window shown in (c) and (d). (g) OA- and LS-trajectories recorded over 5 s show transits of individual beads. (h) Auto-correlation curves of the OA and LS trajectories in (e) reveal interrogation time of ~300 μs. Abbreviations: AE: active element; AcL: acoustic lens; AL: achromatic doublet lens; AMP: low noise amplifier; AWG: arbitrary waveform function generator; DAQ: data acquisition card; GM: galvanometric mirror scanner; IM: inverted microscope; L: planoconvex lens; M: dielectric mirror; ND: neutral density filter; OL: microscope objective lens; PH: pinhole; S: high-precision motorized stage; SP: syringe pump; UT: Ultrasound transducer; W: Waste.*

To test the system's capabilities, we first measured black 10 μm polystyrene bead (μB) as a reference specimen. Fig. 1c and d depict the transit of a single μBs sensed by OA and LS, respectively. In here, the OA signal follows in first approximation a Gaussian curve as expected whereas the LS signal presents a sharp peak occurring at the end of the OA curve. This might be due to the LS detection being tilted and pointing in the same direction as the microfluidic flow, which leads to detecting LS signals at the end of an object's passage. Fig. 1e-f show the corresponding frequency filtered and Hilbert transformed maximum amplitude trajectories. These measurements suggest that the proposed OAFC system enables to measure transits of particles of size ~10 µm. Fig. 1g plots the trajectories of simultaneously recorded OA and LS signals over a record time of 5 s capturing single transit events above threshold, >90% of the events appear in both channels. As depicted in Fig. 1h, both autocorrelations of the OA and LS trajectories exhibit a correlation time of ~300 µs. Theoretically, a flow rate of 100 µL/min through a channel with a cross-section of 100 · 1000 µm² leads to an average flow speed of 16.6 mm/s and a Reynolds number of ~3.03. Measuring beads of 10 µm diameter at the centre of the microfluidic chip, at which the flow speed is expected double the average flow speed based on a laminar Poiseuille flow profile, this should result in an interrogation time of ~300 µs, i.e. ~15 pulses per bead at a repetition rate of 50 kHz at assuming an arbitrary small optical focus, which matches the experimental determined correlation time.

To estimate the performance of the system with specimens of varying size and signal, we next characterized the system using µBs (nominal diameter: 10 µm) and carbon nanoparticles (CNP; size distribution: 2-12 µm) as well as *E. coli* cells expressing chromoprotein labels.

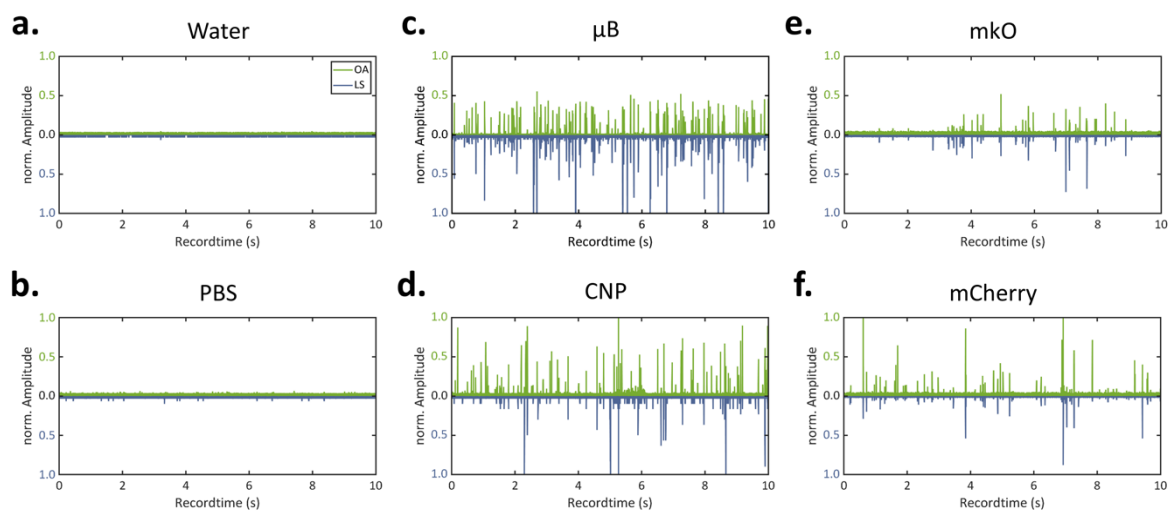

**Fig. 2 Characterization of in-flow signals** at 100 µL/min. Normalized OA and LS trajectories of *(a)* water as the solvent for µB and CNPs as well as of *(b)* PBS as the buffer for the cell suspension exhibit no detected transit event over 10 s record time. Water-suspended *(c)* µB and *(d)* reveal OA and LS signals of transits of individual particles. Normalized OAFC trajectories of E. coli expressing *(e)* mKO or *(f)* mCherry in PBS reveal detectable in-flow signals.

Fig. 2a and b show the background signal level of the aqueous solution used for µB- and CNP-measurements (Fig. 2a) as well as the PBS used for the *E. coli* measurements (Fig. 2b). In both cases, no signals were detected over 10 seconds, which allowed to define a transit-analysis based on thresholds of 400 a.u. for OA and 5 a.u. for LS. Above-threshold OA signals of µBs (Fig. 2d) show a mean value of $1.096*10^3$ +/- 73.1 a.u. and a LS signal of 13.6 +/- 1.24 a.u.. The signal deviations may result from µBs floating through the interrogation area at different positions and, thus, experience different fluence values of the optical excitation cone. For the current system this number defines the maximal achievable precision without signal correction in determining signals from a solution of specimens homogenous in size and signal. Analogous measurements of CNPs indicate an OA signal of 1.806 *10$^3$ +/- 190.7 a.u. and a LS signal of 11.1 +/- 3.19. The increase in spread of OA and LS signal amplitudes is due to the broad size distribution of CNPs of 2-12 µm. As an example of an *in vitro* application we performed similar measurements on mixtures of *E. coli* bacteria expressing either mCherry (UniProtKB: X5DSL3) with an absorption of 72.000 M$^{-1}$cm$^{-1}$ at 587 nm or mKusabira Orange1 (mKO; UniProtKB: Q6I7B2) with 51,600 M$^{-1}$cm$^{-1}$ at 548 nm. Both plots in Fig. 2e and f reveal the system's sensitivity high enough for measuring single-cell transits and sensing both OA of the chromoprotein and LS signals of the cell body. Along with the sensitivity of the system its ability to discriminate different population of cells in a mixture is crucial. Towards this end, we assessed if we are able to distinguish *E. coli* cells expressing the above-mentioned chromophore proteins mCherry and mKO in a mixture in-flow based on their difference in absorption and hence OA signals at the used 532 nm excitation wavelength. Considering both the associated absorption values at 532 nm (i.e. mCherry: 40% -> 28.800 M$^{-1}$cm$^{-1}$; mKO: 58.8 % -> 30.340 M$^{-1}$cm$^{-1}$) as well as their respective quantum yield (i.e. mCherry: 0.22; mKO: 0.6), optoacoustic relevant absorption coefficients can be approximated as 22.000 M$^{-1}$cm$^{-1}$ for mCherry and 12.000 M$^{-1}$cm$^{-1}$ for mKO. Hence, mCherry is expected to generate OA signals being ~1.8 times the ones from mKO, while LS signals from both cultures are expected to resemble each other. We measured pure solution of mCherry or mkO expressing cells as well as a 1:1 mixture of the stock solutions. The pure solutions showed above-threshold peaks of mCherry-cells having an OA signal of 837.5 +/- 153.6 a.u. and a LS signal of 17.3 +/- 4.8 a.u., whereas mKO-cells show an OA signal of 666.5 +/- 91.78 a.u. and a LS signal of 12.1 +/- 5.1 a.u (see Fig. 2e and f). As expected, the LS signal for both cells is comparable whereas the OA signal of cells expressing mCherry is larger than the one of cells expressing mKO. Subtracting the above-mentioned OA background values used for thresholding of 400 a.u., the mean OA signals are rectified to 437.5 a.u. for mCherry and 266.5 a.u. for mKO, thereby approximating a ratio of ~1.65. The large deviations, especially for the OA signal, might be due to differences in expression levels of the cellular population.

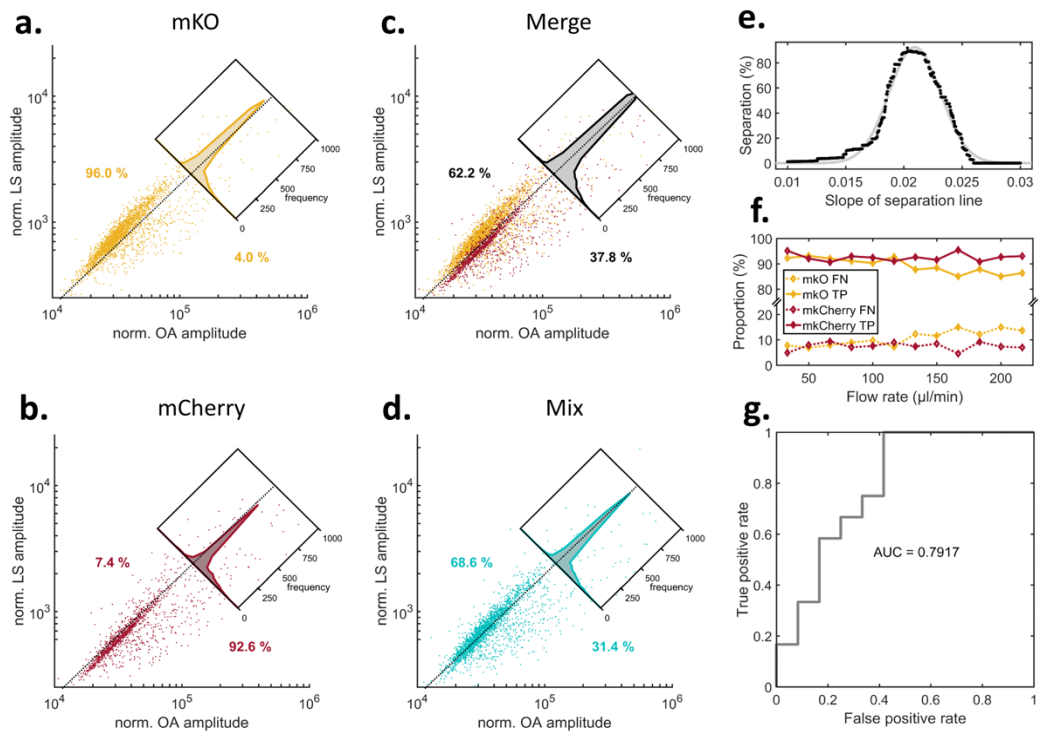

*Fig. 3 Signal distribution of in-flow single transit events of E. coli cells expressing optoacoustic labels enabled separation by applying linear discrimination. (a) Norm. OA signals against norm. LS signals of E. coli cells expressing mKO. Inset depicts separation by linear discrimination of 96.0 %. (b) Analogues analysis of E. coli cells expressing mCherry yield a discrimination of 92.6 %. (c) Data merge of the separate measurements and (d) measurement of cell stock mixture (mixing ratio 1:1) of mKO and mCherry E. coli cells yield a linear discrimination of ~2:1 resembling the ~3:1 concentration difference in stock solutions. (e) Varying the slope of the linear discrimination and Gaussian fitting peaked in ~0.0209 a.u. with an average separation of ~92.5 %. (f) Flow speed analysis shows consistent linear discrimination by constant true-positive (TP) and false-negative (FN) assignments of signals. (g) ROC analysis reveals an AUC of ~0.8.*

Further referencing each OA transit signal to the associated LS signal, a linear relationship between LS and OA signals can be observed (Fig. 3a and b) indicating the validity of correcting OA-signals by LS encoding for the cell-size and relative position in the interrogation area. The insets depict a signal histogram based on an inclined axis, which also enables linear discrimination with a separation precision of >92% for both cases. Further, we merged the two data sets of mKO and mCherry and found a partitioning at ~2:1 (62.2 % vs. 37.8 %) via linear discrimination (Fig. 3c), which was found similarly at ~2:1 (68.6 % vs. 31.4 %) in measuring a 1:1 mixture of the cell stock solutions (Fig. 3d). This resembles roughly the predetermined cell concentrations ratio of the stock solutions being ~3:1 (7.4·$10^8$ cfu/ml for mKO and 2.6·$10^8$ cfu/ml), deviations might be introduced by unequally diluting the stock solutions before measurements. For system optimization, we combined all recorded data and tried to vary the slope of linear discrimination via fitting a Gaussian curve (Fig. 3e), which was found best for 0.0209 a.u. and led to an overall discrimination of ~92.5 %. A separation accuracy of >80 % true-positives (TP) and <20% false-negatives (NG) was consistently achieved at different flow speeds ranging from ~33 to ~233 µl/min (Fig. 3f). We achieved a separation reliability of ~0.8 by area-under-curve (AUC) of a true- vs. false-positive comparison (receiver operator characteristics analysis (ROC), Figure 3g).

**Discussion**

In this work, we demonstrated the proof-of-concept of OAFC: a technique that, based on referencing in-flow optoacoustic signals to simultaneously-recorded light-scattering, allows to characterize the absorbance properties of particles and cells. We showed that the sensitivity of both modalities enabled to detect objects as small as single cells transiting through the system's interrogation area in a microfluidic arrangement. Being independent modalities, the use of LS referencing to increase the reliability of measuring OA signals of single cells in high-throughput flow setups showcased the viability of our OAFC prototype. Future generations of the OAFC will exploit additional signals read along with OA and LS to investigate the objects even more comprehensively. Simultaneously detecting fluorescence emitted by the absorbing entity will complement the accessible information by exploring further optical characteristics (e.g. quantum yield). Further, recording LS signals at more angles, i.e. 45° and 135°, would facilitate a more precise deduction of the objects size.

Exploiting the general OA potential to read intrinsic absorption properties, OA flow cytometers could ultimately use label-free sensing to directly assess a cell's composition of biomolecules[9]. The phenotypical absorption characteristics of a cell are often highly informative regarding the molecular composition, encoding the condition of the cell under investigation, such as the set of expressed proteins[10]. Such characterizations are so far impeded since absorbance measurements of single cells are extremely challenging due to their short absorption path length. Being able to directly measure cellular absorbance characteristics in high-throughput, OAFC concepts geared to investigate molecular specific contrast could enable a wide range of application for life sciences and medical practise to reveal information of a cell that is not addressable via fluorescence labels.

Conclusively, the conception of methods as the here presented OAFC, augmented with sorting capabilities, is an essential step towards both fostering the development of transgene labels to create novel optoacoustic labels using directed-evolution strategies similar to FACS as well as pioneer on label-free microfluidic flow cytometry reducing the preparation time, reagents needed, and cost of conventional methods based on fluorescent.

**Methods**

As schematically depicted in Fig 1a, the OA flow cytometer (OAFC) utilizes the optical excitation of an actively triggered 532 nm laser (SPOT-10-200-532, Elforlight Ltd), which is spatially filtered by a 25 µm pinhole, enlarged by a telescopic arrangement of plan-convex lenses, guided into an inverted microscope (Axio Observer, Zeiss), and focused by a microscopic objective (10 x, 0.45 NA, Zeiss) to a diffraction limited spot. By that, the optical excitation is focused into a microfluidic chip (01-0175-0138-02, microfluidic ChipShop, Jena, Germany) with a flow channel of 1000 µm width and 100 µm height. The channel bottom, through which the optical excitation is focused, consists of a 1 mm thick cyclo-olefine co-polymer, while the top side facing both sensing detectors is made of the same polymer but with only 140 µm thickness. For OA sensing, we equipped a high-frequency ultrasound transducer (HFM23, SONAXIS, France) and for LS sensing an optical fiber (M15L01, Thorlabs) tilted into an 45° angle with respect to the optical axis connected to a high-speed photodiode (DET10A, Thorlabs) equipped with a shortpass optical filter to spectrally block potential fluorescence light. The microfluidic chip is connected to a syringe pump (NE-1000, New Era Pump Systems Inc, USA) in order to push the suspensions with a defined flow speed through the channel. Upon a trigger signal sent out by a function generator (DG1022, Rigol) to the laser as well as the recording data acquisition card (ADQ412, SP Devices), both the amplified (AU-1291, Miteq; 63 dB) OA as well as the LS signal are recorded in a high-speed streaming like process based on temporary buffers. The OA signals are further bandpass filtered in the range of 5-90 MHz. Both signals are then projected using their maximum amplitude of dedicated time windows covering the corresponding signals. The control as well as the subsequent analysis of the yielded flow cytometry trajectories is carried out in Matlab[11–13].

For characterizing and calibrating the OAFC system, we first recorded flowmetry trajectories of suspensions of 10 µm blacked microbeads (µB) and 2-12 µm carbon-nano-particles (CNPs). In all cases, the suspensions were diluted and replenished with ~1 %vol. TWEEN™ 20 (Sigma-Aldrich, Germany) to reduce the surface tension for preventing agglomeration, and with the appropriate amount of sodium-polytungstate (Sigma-Aldrich) for increasing the density of the medium to prevent sedimentation. The required concentration of sodium-polytungstate was determined via calculating the volume fraction of the particles. The prepared suspensions were rested for 1 h to verify proper adjustment of the density. For cell measurements, coding sequences for mKO (UniProtKB: Q6I7B2) and mCherry (UniProtKB: X5DSL3) were synthesized codon optimized for *E. coli* (GeneArt, ThermoFisher) and inserted in pET21a(+) vector (Novagene) using the NdeI and XhoI (ThermoFisher) restriction sites. Proteins were expressed in BL21 DE3 (New England Biolabs) at 37 °C and 180 rpm after Isipropyl-thiogalactoside (IPTG) induction. Cell density and protein expression was monitored by optical absorption measurements at 600 nm and fluorescence measurements at the respective peak absorption wavelengths of the two proteins.

**Acknowledgements**

M.S. acknowledge funding from the German Research Foundation (DFG) grants "Gottfried Wilhelm Leibniz Prize 2013" (NT 3/10–1), CRC 1123 (Z1) and the Reinhart Koselleck award "High resolution near-field thermoacoustic sensing and imaging" (NT 3/9–1). A.C.S. and the conceptualization of OAFC received funding from the Deutsche Forschungsgemeinschaft (STI656/1-1).